\documentclass[aps,floats]{revtex4}
\usepackage{amsmath,amssymb}
\usepackage{graphicx,epsfig}
\usepackage[greek,english]{babel}

\begin{document}
\bibliographystyle {plain}

\def\oppropto{\mathop{\propto}} 
\def\opsimeq{\mathop{\simeq}}
\def\opoverderline{\mathop{\overline}}
\def\operarrow{\mathop{\longrightarrow}}
\def\opsim{\mathop{\sim}}

\def\fig#1#2{\includegraphics[height=#1]{#2}}
\def\figx#1#2{\includegraphics[width=#1]{#2}}


\title{ Multifractality in the generalized Aubry-Andr\'e quasiperiodic localization model \\
with power-law hoppings or power-law Fourier coefficients    } 


\author{ C\'ecile Monthus }
 \affiliation{Institut de Physique Th\'{e}orique, 
Universit\'e Paris Saclay, CNRS, CEA,
91191 Gif-sur-Yvette, France}

\begin{abstract}
The nearest-neighbor Aubry-Andr\'e quasiperiodic localization model is generalized to include power-law translation-invariant hoppings $T_l\propto t/l^a$ or power-law Fourier coefficients $W_m \propto w/m^b$ in the quasi-periodic potential. The Aubry-Andr\'e duality between $T_l$ and $W_m$ is manifest when the Hamiltonian is written in the real-space basis and in the Fourier basis on a finite ring. The perturbative analysis in the amplitude $t$ of the hoppings yields that the eigenstates remain power-law localized in real space for $a>1$ and are critical for $a_c=1$ where they follow the Strong Multifractality linear spectrum, as in the equivalent model with random disorder.  The perturbative analysis in the amplitude $w$ of the quasi-periodic potential yields that the eigenstates remain delocalized in real space (power-law localized in Fourier space) for $b>1$ and are critical for $b_c=1$ where they follow the Weak Multifractality gaussian spectrum in real space (or Strong Multifractality linear spectrum in the Fourier basis). This critical case $b_c=1$ for the Fourier coefficients $W_m$ corresponds to a periodic function with discontinuities, instead of the cosinus of the standard self-dual Aubry-Andr\'e model. 

\end{abstract}

\maketitle

\section{ Introduction }

Although the phenomenon of Anderson Localization is mostly studied in the presence of random potentials (see the reviews \cite{50years,janssenrevue,mirlin_revue2000,mirlin_revue}),
the case of non-random quasi-periodic potentials has also attracted a lot of interest over the years
\cite{harper,hof,azbel,aubry,thou,pietro,abe,has,han,piechon,pichard,aulbach,leboeuf,igloi,sarma,chandran,dhar,znidaric}.
More recently in the presence of interactions, the Many-Body-Localization (see the reviews \cite{revue_huse,revue_altman}
and references therein) has been also considered for the quasi-periodic case \cite{refael,alt,huse,nag,lee} in order to understand the similarities and the differences with the random case.

In the simplest case of the nearest-neighbor one-dimensional Anderson Localization model, it is well known
 that any random potential (even extremely weak) leads to exponentially localized eigenfunctions,
while the corresponding nearest-neighbor self-dual Aubry-Andr\'e quasiperiodic model \cite{aubry} displays
 a phase transition between a localized phase and a delocalized ballistic phase.
In the presence of power-law long-ranged hoppings $T(l) \propto l^{-a}$, one-dimensional Anderson localization models with disorder
are know to become critical for the value $a_c=1$ where the multifractality of eigenfunctions changes continuously as a function of the amplitude between strong and weak multifractality (see the review \cite{mirlin_revue} and references therein).
In the present paper, we thus wish to consider similarly the long-ranged version of the 
Aubry-Andr\'e quasiperiodic model, where the hoppings decay only as a power-law $T(l) \propto l^{-a}$.
As a consequence of the Aubry-Andr\'e duality \cite{aubry}, it will be also interesting to focus on the case 
where the Fourier coefficients of the quasi-periodic potential decay only as a power-law $W_m \propto m^{-b}$.
The main goal is to analyze the multifractal properties of eigenstates in these two cases.

The paper is organized as follows.
In section \ref{sec_inf}, the generalized version of the Aubry-Andr\'e model with arbitrary translation-invariant hoppings 
and arbitrary Fourier coefficients
is defined on the infinite lattice.
In section \ref{sec_ring}, we describe the properties of the model for a finite ring of $N$ sites, where the Aubry-Andr\'e
duality between the $N$ sites and the  $N$ Fourier modes is obvious.
In section \ref{sec_strong}, the case of power-law hoppings with a small amplitude is studied perturbatively
and leads to the strong multifractality spectrum of eigenstates.
In section \ref{sec_weak}, the case of power-law Fourier coefficients with a small amplitude 
is studied perturbatively and leads to the weak multifractality spectrum.
Our conclusions are summarized in \ref{sec_conclusion}.
In Appendix \ref{sec_ran}, some differences with the random case are stressed.

\section{ Generalized Aubry-Andr\'e on the infinite lattice }

\label{sec_inf}

In this paper, we consider the generalized version of the one-dimensional Aubry-Andr\'e Hamiltonian \cite{aubry}
containing two 
competing contributions
\begin{eqnarray}
H && = H^{deloc} + H^{loc} 
\label{H}
\end{eqnarray}
where $ H^{deloc}$ contains translation-invariant hoppings that may be long-ranged,
and where $H^{loc}$ contains on-sites energies following a quasi-periodic function containing arbitrary Fourier coefficients.
Let us now discuss separately their properties.

\subsection{ $H^{deloc}$ containing translation-invariant hoppings }

The contribution $H^{deloc} $ contains hoppings $H^{deloc}_{nn'}$ between different sites $n \ne n'$,
with the translation-invariance property
\begin{eqnarray}
H^{deloc}_{nn'}=T_{n-n'}
\label{transla}
\end{eqnarray}
and the hermitian property $H^{deloc}_{nn'}=(H^{deloc}_{n'n})^*$ corresponding to 
\begin{eqnarray}
T_l=T_l^*
\label{translastar}
\end{eqnarray}

On the infinite lattice, $H^{deloc}  $ can be then rewritten with the various forms
\begin{eqnarray}
H^{deloc} && = \sum_{n \ne n'} H_{nn'} \vert n> < n' \vert
=  \sum_{n \ne n'} T_{n-n'} \vert n> < n' \vert
\nonumber \\
&& =  \sum_{n=-\infty}^{+\infty} \sum_{l \ne 0} T_l  \vert n> < n-l \vert
=  \sum_{n=-\infty}^{+\infty} \sum_{l=1}^{+\infty} 
  ( T_l \vert n> < n-l \vert+T^*_l \vert n-l> < n \vert )
\label{Hdeloc}
\end{eqnarray}
The translation invariance yields that its eigenstates are delocalized Fourier modes.

\subsection{ $H^{loc}$ containing quasi-periodic on-site energies }

The contribution 
\begin{eqnarray}
 H^{loc} && = \sum_{n=-\infty}^{+\infty} H_{nn} \vert n> < n \vert
\label{Hloc}
\end{eqnarray}
contains on-site energies $H_{nn}$ following the quasi-periodic form
\begin{eqnarray}
H_{nn}  =   W( g n ) 
\label{quasiperiogene}
\end{eqnarray}
where $g$ is usually chosen in terms of the inverse Golden mean
\begin{eqnarray}
 \frac{g}{2 \pi} = \frac{\sqrt{5}-1}{2}  
\label{golden}
\end{eqnarray}
while $W(x)$ is a real $2 \pi$-periodic function of zero-mean 
that can be defined by its Fourier expansion
\begin{eqnarray}
 W(x) = \sum_{m=-\infty}^{+\infty} W_m e^{i mx} =
\sum_{m=1}^{+\infty} (W_m e^{i mx} + W_m^*e^{-imx})
\label{wxf}
\end{eqnarray}
where the Fourier coefficients satisfy $W_{-m}=W_m^*$ and $W_{m=0}=0$.

\subsection{ Standard nearest-neighbor self-dual Aubry-Andr\'e  model}

The standard self-dual Aubry-Andr\'e  model \cite{aubry} corresponds
to hoppings limited to nearest-neighbors $l=\pm 1$
\begin{eqnarray}
T^{(AA)}_l= \delta_{l,1}+\delta_{l,-1}
\label{tlAA}
\end{eqnarray}
and to the function $W(x)= \lambda 2 \cos (x+h)$ involving only the first Fourier coefficients $m = \pm 1$
\begin{eqnarray}
W^{(AA)}_m= \lambda e^{imh}  (  \delta_{m,1}+\delta_{m,-1})
\label{WmAA}
\end{eqnarray}
Aubry and Andr\'e \cite{aubry} have shown that this model satisfy a remarkable self-duality.
The main consequence is that the exact critical point $\lambda_c=1$ separates the delocalized ballistic phase $\lambda<1$
from the localized phase $\lambda>1$ characterized by the exact localization length $\xi = \frac{1}{\ln \lambda}$ \cite{aubry}.

\subsection{ Long-Ranged power-law hoppings}

In the present paper, we wish to analyze the cases where the hoppings decay as a power-law with some exponent $a \geq 1$
\begin{eqnarray}
T^{(a)}_l = i t \frac{ (1-\delta_{l,0} ) {\rm sgn}(l) }{ \vert l \vert^a}
\label{ta}
\end{eqnarray}
including the critical case $a_c=1$
\begin{eqnarray}
T^{(a_c=1)}_l = it \frac{ (1-\delta_{l,0} )}{l}
\label{tcriti}
\end{eqnarray}
known as the Calogero-Moser matrix model (see \cite{ogweak} and references therein).
In particular we will focus in section \ref{sec_strong}
on the strong multifractality regime occuring for the critical case $a_c=1$ with a small amplitude $t$.

Note that the presence of the imaginary factor $i$ in the critical case of Eq \ref{tcriti}
is essential to obtain a well defined model for $a_c=1$ : the real symmetric power-law case $T(l)= \frac{t}{\vert l \vert^a}$ 
which has been studied in various contexts \cite{rodriguez,moura,buy,moessner}
is well defined only for $a>1$, because the Fourier mode $K=0$ would have an infinite energy for $a_c=1$.
For $a>1$, there is no Anderson transition in the middle of the spectrum,
but only near the edge of the spectrum where the level spacing is anomalous 
(see section \ref{sec_rod} in Appendix).

\subsection{ Long-Ranged Fourier coefficients }

Since there exists some duality between the hoppings $T_l$ and the Fourier coefficients $W_m$ (see more details in section \ref{sec_duality}),
we will also focus on the case where the Fourier coefficients $W_m$ decay as a power-law with some exponent exponent $b \geq 1$
\begin{eqnarray}
W^{(b)}_m =-iw \frac{(1-\delta_{m,0} ) {\rm sgn} (m) }{ \vert m \vert^b}
\label{wb}
\end{eqnarray}
including the critical case $b_c=1$
\begin{eqnarray}
W^{(b_c=1)}_m=-i w \frac{ (1-\delta_{m,0} )}{m}
\label{wcriti}
\end{eqnarray}
where the weak multifractality regime occuring for small amplitude $w$ will be analyzed in section \ref{sec_weak}.

From the general theory of Fourier series, it is well-known that 
the decay of the Fourier coefficients $W_m$ for large $m$
directly reflects the regularity properties of the function $W(x)$ of Eq. \ref{wxf} :
the critical decay as $1/m$ corresponds to a function $W(x)$ presenting discontinuities,
the decay as $1/m^2$ corresponds to a continuous function $W(x)$ whose first derivative presents discontinuities, and so on.

For instance, the critical case $b_c=1$ of Eq. \ref{wcriti} corresponds to the $2\pi$-periodic linear function
\begin{eqnarray}
W^{(b_c=1) (N=\infty)}(0<x<2 \pi) && = 2 w  \sum_{m=1}^{+\infty} \frac{1}{m}   \sin \left( mx  \right) =  w (\pi-x) \ \ \ { \rm for } \ \  0<x<2 \pi
\label{wramp}
\end{eqnarray}
 with the following discontinuity at $x=0 [{\rm modulo } 2 \pi]$
\begin{eqnarray}
W^{(b_c=1) (N=\infty)}(x=0^+) && = w \pi
\nonumber \\
W^{(b_c=1) (N=\infty)}(x=0) && = 0
\nonumber \\
W^{(b_c=1) (N=\infty)}(x=0^-) && = - w \pi
\label{wdiscontinuum}
\end{eqnarray}

\section{  Generalized version of the Aubry-Andr\'e model on a finite ring of $N$ sites }

\label{sec_ring}

Since our goal is to analyze the multifractal properties of critical eigenstates via perturbation theory, 
it is convenient to consider the finite-size version of the above model in order to have discrete energy levels.
Another advantage is that the Aubry-Andr\'e duality is much clearer when the $N$ sites and the $N$ Fourier modes 
play exactly similar roles, even if the duality has been first formulated for the infinite lattice \cite{aubry}.
In this section, we thus 
describe how the infinite-lattice model described in the previous section can be defined on a finite ring containing 
$N$ sites with periodic boundary conditions  $\vert n+N>=\vert n>$.
To avoid discussions on the differences between even and odd sizes $N$,
it will be convenient to focus only on the odd case with the notation
\begin{eqnarray}
N=2P+1
\label{odd}
\end{eqnarray}

\subsection{ Fourier diagonalization of $H^{deloc}$ }

It is convenient to define the finite-size version of Eq. \ref{Hdeloc} as
\begin{eqnarray}
H^{deloc(N)}
&& =  \sum_{n=0}^{N-1} \sum_{l=1}^{P} 
  ( T_l \vert n> < n-l \vert+T^*_l \vert n-l> < n \vert )
\label{Hdelocn}
\end{eqnarray}

In the basis $K=0,..,N-1$ of the $N$ Fourier modes
\begin{eqnarray}
\vert K > = \frac{1}{\sqrt N} \sum_{n=0}^{N-1} e^{i 2 \pi K \frac{n}{N}} \vert n>
\label{psifourier}
\end{eqnarray}
with the orthonormalization
\begin{eqnarray}
< K' \vert K > = \frac{1}{ N} \sum_{n=0}^{N-1} e^{i 2 \pi (K-K') \frac{n}{N}} = \delta_{K',K}
\label{psifourierorthog}
\end{eqnarray}
one obtains the diagonal form
\begin{eqnarray}
H^{deloc(N)} && =  \sum_{K=0}^{N-1}    E^{deloc}_{K}\vert K> <K\vert
\label{Hdelockkp}
\end{eqnarray}
with the Fourier eigenvalues
\begin{eqnarray}
E^{deloc}_{K} && =  \sum_{l=1}^{P} (T_l e^{-i 2 \pi K \frac{l}{N}} +T^*_l e^{i 2 \pi K \frac{l}{N}} )
\label{edelocK}
\end{eqnarray}

For instance the nearest-neighbor hopping of Eq. \ref{tlAA} corresponds to the usual result
\begin{eqnarray}
E^{deloc(AA)}_{K} && = 2 \cos \left( 2 \pi K \frac{l}{N} \right)
\label{edelAA}
\end{eqnarray}
while the power-law case of Eq. \ref{ta} corresponds to
\begin{eqnarray}
E^{deloc(a)}_{K} && = 2 t  \sum_{l=1}^{P} \frac{1}{l^a}   \sin \left( 2 \pi K \frac{l}{N} \right)
\label{Hkkca}
\end{eqnarray}
In the thermodynamic limit $N \to +\infty$ where the momentum $k= 2 \pi \frac{K}{N} $ becomes continuous in the Brillouin zone $[0,2 \pi[$,
the energy
\begin{eqnarray}
E^{deloc(a)(N=\infty)}(k) && = 2 t  \sum_{l=1}^{+\infty} \frac{1}{l^a}   \sin \left( kl  \right) 
\label{econtinuum}
\end{eqnarray}
remains finite for $a > 1$ where the Fourier series converges absolutely.
For the critical case $a_c=1$ of Eq. \ref{tcriti}, the absolute convergence is lost,
but one recognizes
 the sine-Fourier decomposition of the odd $2 \pi$-periodic function of Eq. \ref{wramp} following the linear form on $]0,2 \pi[$
\begin{eqnarray}
E^{deloc(a_c=1) (N=\infty)}(0<k<2 \pi) && = 2 t  \sum_{l=1}^{+\infty} \frac{1}{l}   \sin \left( kl  \right) =  t (\pi-k) \ \ \ { \rm for } \ \  0<k<2 \pi
\label{ramp}
\end{eqnarray}
 with the following discontinuity at $k=0 [{\rm modulo } 2 \pi]$
\begin{eqnarray}
E^{deloc(a_c=1) (N=\infty)}(k=0^+) && = t \pi
\nonumber \\
E^{deloc(a_c=1) (N=\infty)}(k=0) && = 0
\nonumber \\
E^{deloc(a_c=1) (N=\infty)}(k=0^-) && = - t \pi
\label{discontinuum}
\end{eqnarray}
For large $N$,  the Fourier modes $K=1,..,N-1$ thus follow the linear ramp of Eq. \ref{econtinuum}
\begin{eqnarray}
E^{deloc(a_c=1)}_{K} && \simeq  t  \left( \pi- 2 \pi  \frac{K}{N}  \right) \ \ \ {\rm for } \ \ K=1,..,N-1
\label{edelocdiscrete}
\end{eqnarray}
while the mode $K=0$ is exactly in the middle of the spectrum
\begin{eqnarray}
E^{deloc(a_c=1)}_{K=0} && = 0
\label{edeloczero}
\end{eqnarray}
and has for neighboring energy levels the Fourier modes corresponding to $K=P$ et $K=P+1$
\begin{eqnarray}
E^{deloc(a_c=1}_{K=P} && =  \frac{\pi t}{N}
\nonumber \\
E^{deloc(a_c=1}_{K=P+1} && = - \frac{\pi t}{N}
\label{edelocnearzero}
\end{eqnarray}

\subsection{ Properties of the quasiperiodic on-site energies $H_{nn}$ on a finite ring }

On a finite ring of $N$ sites,  it is convenient to keep only the Fourier modes $-P \leq m \leq P$
of the $2 \pi$ periodic function of Eq. \ref{wxf}
\begin{eqnarray}
 W_N(x) = \sum_{m=-P}^{+P} W_m e^{i m x } = 
\sum_{m=1}^{P} (W_m e^{i mx} + W_m^*e^{-imx})
\label{qfinite}
\end{eqnarray}
and to ask that the quasi-periodic form of Eq. \ref{quasiperiogene}
\begin{eqnarray}
H_{nn}  =   W_N( g n ) = \sum_{m=-P}^{+P} W_m e^{i m g n } 
\label{quasiperiogenen}
\end{eqnarray}
is compatible with the periodic boundary condition $n \to n+N$ of the ring.
This condition 
$e^{i g N}=1$ yields the choice
\begin{eqnarray}
 \frac{g}{2 \pi} = \frac{G}{N}  
\label{qfiniten}
\end{eqnarray}
where the integer $G=G_N$ is chosen to ensure that the successive 
fractions $  \frac{G_N}{N}  $ converge towards the inverse Golden mean Eq. \ref{golden} :
the standard solution consists in choosing $N=F_i$ and $G_N=F_{i-1}$
in terms of the Fibonacci numbers satisfying the recurrence $F_i=F_{i-1}+F_{i-2}$.

In summary, the on-site energies  on the finite ring read
\begin{eqnarray}
H_{nn}  = \sum_{m=-P}^{+P}  W_m e^{i m 2 \pi  \frac{G}{N}  n}
= \sum_{m=1}^P \left(  W_m e^{i m 2 \pi  \frac{G}{N}  n} + W_m^* e^{- i m 2 \pi  \frac{G}{N}  n}\right)
\label{epsfquasi}
\end{eqnarray}

In terms of the Fourier modes of Eq. \ref{psifourier}, the localized contribution of the Hamiltonian reads
\begin{eqnarray}
H^{loc} && = \sum_{n=0}^{N-1} H_{nn} \vert n> < n \vert
=  \sum_{K=0}^{N-1}  \sum_{K'=0}^{N-1}  H^{loc}_{KK'} \vert K> < K' \vert
\label{Hkkp}
\end{eqnarray}
with
\begin{eqnarray}
H^{loc}_{KK'} && = \frac{1}{N} \sum_{n=0}^{N-1} e^{ i 2 \pi (K'-K) \frac{n}{N}}  H_{nn} 
= \sum_{m=-P}^{+P}  W_m  \frac{1}{N} \sum_{n=0}^{N-1} e^{ i 2 \pi (K'-K+mG) \frac{n}{N}} 
\nonumber \\
&& = \sum_{m=-P}^{+P}  W_m   \delta_{K'-K+mG [N] }
\label{Hlockkp}
\end{eqnarray}
where the notation $[N]$ means $( {\rm modulo} \ \ N )$.
So the interaction between two Fourier modes $K$ and $K'$ depend on the integer $m$
satisfying $K-K' = m G [N ]$.
For instance, in the Aubry-Andr\'e case of Eq.  \ref{WmAA},
Eq. \ref{Hlockkp} reduces to
\begin{eqnarray}
 H^{loc}_{KK'} && = \lambda e^{ih}   \delta_{K'-K+G [N] }
+ \lambda e^{-ih}   \delta_{K'-K-G [N] }
\label{epsfquasinn}
\end{eqnarray}
so that each Fourier mode $K$ interacts only with two other Fourier modes $K'=K \pm G [N]$.

\subsection{ Quasiperiodic Fourier basis  }

The form of Eq. \ref{Hlockkp} for the interaction between two Fourier modes $(K,K')$ suggests 
that it is useful to reparametrize the Fourier modes $K=0,..,N-1$ 
in terms of the new integer $Q=0,..,N-1$ satisfying \cite{aulbach,igloi}
\begin{eqnarray}
K= Q G [N]
\label{KQ}
\end{eqnarray}
This amounts to introduce the alternative Fourier basis adapted to the quasi-periodicity
\begin{eqnarray}
\vert Q > = \frac{1}{\sqrt N} \sum_{n=0}^{N-1} e^{i 2 \pi Q \frac{G n}{N}} \vert n>
\label{Qfourier}
\end{eqnarray}
The orthonormalization
\begin{eqnarray}
< Q' \vert Q > = \frac{1}{ N} \sum_{n=0}^{N-1} e^{i 2 \pi (Q-Q') \frac{G n}{N}} = 
\delta_{(Q-Q')G=0 [N]} =
\delta_{Q',Q}
\label{psifourierorthogq}
\end{eqnarray}
can be understood as follows : besides the trivial solution $Q=Q'$,
one cannot find another solution $(Q-Q')G=jN  $ corresponding to some integer $j \ne 0$ :
indeed the equation $\frac{j}{ Q-Q'}=\frac{G}{N }  $ has no solution for
  $Q-Q'=1,..,N-1$ since the irreducible fraction $ \frac{G}{N } $ cannot be written as a fraction with a smaller denominator.

The advantage of this new basis is that the interaction of Eq. \ref{epsfquasi} becomes much simpler
\begin{eqnarray}
H^{loc}_{QQ'} 
&& = \sum_{m=-P}^{+P}  W_m   \delta_{(Q'-Q+m)G [N] } = W_{m=Q-Q' [N]}
\label{Hlockkq}
\end{eqnarray}
as it is directly given by the Fourier coefficient $W_m$ corresponding to $m=Q-Q' [N]$.
For instance, in the Aubry-Andr\'e case of Eq.  \ref{WmAA},
the interaction is only between nearest-neighbors for the new indices $(Q,Q')$
\begin{eqnarray}
H^{loc}_{Q Q'} && =
   W_1 ( e^{i  h } \delta_{Q' -Q+1 } +  e^{- i  h } \delta_{Q' -Q-1 }  )
\label{qqpnn}
\end{eqnarray}

The other contribution $ H^{deloc}$ that was diagonal in $K$ remains of course diagonal in $Q$,
but the corresponding eigenvalues $E^{deloc}_K$ of Eq. \ref{edelocK} that were well ordered in $K$ become
\begin{eqnarray}
E^{deloc}_{Q} && =  \sum_{l=1}^{P} (T(l) e^{-i 2 \pi Q G \frac{l}{N}} +T^*(l) e^{i 2 \pi QG  \frac{l}{N}} )
\label{Hdelocqq}
\end{eqnarray}
This is thus similar to the quasi-periodic form of the on-site energies of Eq. \ref{epsfquasi}
where $W_m$ has been replaced by $T^*(l)$.

\subsection{ Duality between the real-space basis and the quasiperiodic Fourier basis  }

\label{sec_duality}

The discussion of the quasiperiodic Fourier basis of the previous section
shows the following Aubry-Andr\'e duality for the finite-size model that we consider :

(i) In the real space basis $\vert n>$, the delocalized Hamiltonian $H^{deloc}$ is a circulant matrix
defined by the coefficients $T_l$
\begin{eqnarray}
H^{deloc}_{nn'}=T_{n-n'}
\label{circulent}
\end{eqnarray}
while the localized Hamiltonian $H^{loc}$ is diagonal and follows the quasiperiodic form defined by Fourier coefficients $W_m$
\begin{eqnarray}
H^{loc}_{nn}  = \sum_{m=-P}^{+P}  W_m e^{i m 2 \pi  \frac{G}{N}  n}
= \sum_{m=1}^P \left(  W_m e^{i m 2 \pi  \frac{G}{N}  n} + W_m^* e^{- i m 2 \pi  \frac{G}{N}  n}\right)
\label{hnnwm}
\end{eqnarray}

(ii) In the quasiperiodic Fourier basis $\vert Q>$, the delocalized Hamiltonian $H^{deloc}$ is 
diagonal and follows the quasiperiodic form of Fourier coefficients $T_l^*$
\begin{eqnarray}
E^{deloc}_{QQ} && =  \sum_{l=1}^{P} (T(l) e^{-i 2 \pi Q G \frac{l}{N}} +T^*(l) e^{i 2 \pi QG  \frac{l}{N}} )
\label{Hqq}
\end{eqnarray}
while the localized Hamiltonian $H^{loc}$ is a circulant matrix
defined by the coefficients$W_m$
\begin{eqnarray}
H^{loc}_{QQ'} 
&& = W_{Q-Q' }
\label{Hkkq}
\end{eqnarray}

For instance, the standard Aubry-Andr\'e duality corresponds to the case where both $T_l$ and $W_m$
are limited to nearest neighbors
\begin{eqnarray}
T^{(AA)}_l= \delta_{l,1}+\delta_{l,-1}
\nonumber 
W^{(AA)}_m= \lambda   (  \delta_{m,1}+\delta_{m,-1})
\label{WmAAhzero}
\end{eqnarray}
so that the self-dual point $\lambda_c=1$ correspond to the critical point between the localized and the delocalized phases.

For the present power-law models that we consider, this Aubry-Andr\'e duality
relates the power-law hopping case of Eq. \ref{ta} and the power-law Fourier coefficients of Eq. \ref{wb}.

\section{ Perturbation theory in the amplitude $t$ of the power-law hoppings }

\label{sec_strong}

In this section, we focus on the case of Eq \ref{ta} where the hoppings $T_l$ decay as a power-law of exponent $a \geq 1$
with some small amplitude $t$,
while the Fourier coefficients $W_m$ of finite amplitude can be either short-ranged or long-ranged.

\subsection{ Perturbation theory for the eigenstates }

If $H^{deloc}=0$, the $N$  eigenstates are completely localized on the sites $n$
\begin{eqnarray}
\vert \psi^{loc(0)}_{n} > && = \vert n >
\label{psiloc0}
\end{eqnarray}
and the corresponding eigenvalues given by the on-site energies
\begin{eqnarray}
E^{loc(0)}_{n}  && = H_{nn}
\label{eloc0}
\end{eqnarray}
are non-degenerate.
The first order perturbation theory in $H^{deloc}$ yields the eigenstates 
\begin{eqnarray}
\vert \psi^{loc(0+1)}_{n} >&& = \vert n > + \sum_{n'} \vert n' >  \frac{ H^{deloc}_{n'n} }{H_{nn}-H_{n'n'}} 
\nonumber \\
&& = \vert n > +\sum_{l=-P,+P} 
 \vert n+l > \frac{ T_l }
{H_{nn}-H_{n+l,n+l}} 
\label{perst}
\end{eqnarray}

For a given eigenstate indexed by the site $n$ of the zero-order,
the density on the other sites $l=1,..,N-1$ is thus given at lowest order by
\begin{eqnarray}
\rho_l \equiv \left \vert  <n+l \vert  \psi^{loc(0+1)}_{n} > \right \vert^2
= \left \vert  \frac{ T_l }
{H_{nn}-H_{n+l,n+l}} \right \vert^2
\label{pl}
\end{eqnarray}
In this perturbation theory, one expects that the most dangerous term corresponds
to the smallest denominator associated to the closest energy level :
the difference $(H_{nn}-H_{n+l,n+l}) $ between these two neighboring energy levels scales as the level spacing 
\begin{eqnarray}
\Delta_N \propto \frac{1}{N}
\label{levelspacing}
\end{eqnarray}
This has to be compared with the scaling of the hopping on the scale of the system-size $l \propto N$
\begin{eqnarray}
T_{l } \oppropto_{l \propto N} \frac{1}{N^a}
\label{tlmax}
\end{eqnarray}
This scaling argument yields that the perturbation theory remains consistent at large size $N$ only for $a>1$,
while $a_c=1$ corresponds to the critical case. 

\subsection{ Multifractal analysis }

In the multifractal formalism, one is interested into the singularity spectrum $f(\alpha)$ governing the leading exponential 
behavior of the probability that the density of Eq. \ref{pl} scales as $\rho_l \propto N^{-\alpha}$ \cite{mirlin_revue}
\begin{eqnarray}
{ \cal P}( \alpha ) \propto N^{f(\alpha)-1}
\label{falpha}
\end{eqnarray}

For the energy difference $ \vert H_{nn}-H_{n+l,n+l} \vert $ appearing in the denominator of Eq. \ref{pl}, 
the important property is the level spacing of Eq. \ref{levelspacing}.
For instance, there are a finite number $O(1)$ of states that have an energy difference scaling as the level spacing $\frac{1}{N} $,
while there is an extensive number $O(N)$ of states that have a finite energy difference $O(1)$.
More generally, if the states are re-labelled in the order of the energy difference with some index $p=1,..,N$,
the change of variables $p=N^x$ with $0 \leq x \leq 1$ and $dp =N^x (\ln N) dx $
yields that the number of states having an energy difference of order $N^x (\Delta_N)=N^{x-1}$ scales as $N^x (\ln N)  $.
So the probability to have an energy difference of order $N^{x-1}$ scales as
\begin{eqnarray}
{\rm Prob} ( \vert H_{nn}-H_{n+l,n+l} \vert \propto N^{x-1} ) \propto N^{x-1} (\ln N) \theta(0 \leq x \leq 1)
\label{mx}
\end{eqnarray}

For the power-law hopping $T_l$ of Eq. \ref{ta} appearing in the numerator of Eq. \ref{pl},
the change of variable $l=N^y$ with $0\leq y \leq 1$ and $dl = N^y (\ln N) dy$ yields that 
the probability to have a hopping scaling as $\vert T_l \vert \propto N^{-a y}$ scales as
\begin{eqnarray}
{\rm Prob} ( \vert T_l \vert \propto N^{-ay} ) \propto N^{y-1} (\ln N) \theta(0 \leq y \leq 1)
\label{my}
\end{eqnarray}

So the probability of Eq. \ref{falpha} to have a density $\rho_l $ of Eq. \ref{pl} scaling as $N^{-\alpha}$
can be evaluated from Eqs \ref{mx} and \ref{my} with the correspondence $N^{-\alpha}= \left( \frac{N^{-ay} }{N^{x-1}} \right)^2$
as
\begin{eqnarray}
{ \cal P}_a( \alpha ) && \propto \int_0^1 dx (\ln N) N^{x-1} \int_0^1 dy (\ln N) N^{y-1} \delta ( \alpha +2-2x-2 ay)
\nonumber \\
&& = (\ln N)^2 N^{\frac{\alpha}{2}-1 } \int_0^1 dy   N^{(1 -a)y} \ \ \theta\left ( \frac{\alpha}{2a}  \leq y \leq \frac{\alpha}{2a}+\frac{1}{a} \right)
\label{faalpha}
\end{eqnarray}
Since the exponent $\alpha$ is positive $\alpha \geq 0$, the lower bound of the integral is $y_{min}=\frac{\alpha}{2a} \geq 0  $.

For the critical case $a_c=1$, Eq. \ref{faalpha} becomes
\begin{eqnarray}
{ \cal P}_{a_c=1}( \alpha ) 
&& = (\ln N)^2 N^{\frac{\alpha}{2}-1 } \int  dy  \ \ \theta\left ( \frac{\alpha}{2}  \leq y \leq 1 \right)
\nonumber \\
&& = \left (1- \frac{\alpha}{2}   \right)  (\ln N)^2 N^{\frac{\alpha}{2}-1 } \theta (0 \leq \alpha \leq 2)
\label{facalpha}
\end{eqnarray}
so that the corresponding multifractal spectrum of Eq. \ref{falpha}
\begin{eqnarray}
f_{a_c=1}(\alpha) = \frac{\alpha}{2} \theta ( 0 \leq \alpha  \leq 2  )
\label{fac}
\end{eqnarray}
is the well-known  'Strong Multifractality' critical spectrum \cite{mirlin_fyodorov,mirlin_four} found at Anderson Localization Transition
in the limit of infinite dimension $d \to +\infty$ \cite{mirlin_revue}.
It appears similarly in the long-ranged power-law hopping model in the presence of random (instead of quasi-periodic) on-site energies
 \cite{levitov1,levitov2,levitov3,levitov4,mirlin_evers,fyodorov,
fyodorovrigorous,oleg1,oleg2,oleg3,oleg4,oleg5,oleg6,ogweak,olivier_conjecture,us_strongmultif}.
It has been also found in various matrix models, in particular
 in the generalized Rosenzweig-Potter matrix model of \cite{kravtsov_rosen}, and in the L\'evy Matrix Model \cite{c_levyloc},
as well as in Many-Body-Localization models \cite{c_mblrgeigen,c_mblentropy}.

For $a > 1$, the integral over $y$ in Eq. \ref{faalpha}
is dominated 
by the lower value $y_{min}=\frac{\alpha}{2a}   $, so that it is convenient to make
 the change of variables $y=\frac{\alpha}{2a}   + \frac{u}{\ln N}$ to obtain
\begin{eqnarray}
{ \cal P}_{a>1}( \alpha ) 
&& = (\ln N)^2 N^{\frac{\alpha}{2}-1 } \int_0^{(1-\frac{\alpha}{2a}  ) \ln N }   \frac{du}{\ln N}   N^{(1 -a) \frac{\alpha}{2a}  + (1-a) \frac{u}{\ln N}} \ \ 
\theta\left ( 0 \leq \alpha \leq 2a  \right)
\nonumber \\
&& =  ( \ln N )  N^{\frac{\alpha}{2a }-1 } \theta\left ( 0 \leq \alpha \leq 2a  \right)
\int_0^{(1-\frac{\alpha}{2a}  ) \ln N }   du  e^{ -(a-1) u } \ \ 
\nonumber \\
&& =  ( \ln N ) N^{\frac{\alpha}{2a }-1 } \theta\left ( 0 \leq \alpha \leq 2a  \right)
\frac{ 1 -   N^{- (a-1) (1-\frac{\alpha}{2 a } ) }}{a-1} 
\label{fabalpha}
\end{eqnarray}
so that the corresponding multifractal spectrum of Eq. \ref{falpha}
\begin{eqnarray}
f_{a>1}(\alpha) = \frac{\alpha}{2 a} \theta ( 0 \leq \alpha  \leq 2 a  )
\label{fab}
\end{eqnarray}
is also linear as Eq. \ref{fac} and has been been found in other models 
in the localized phase close to the critical point describe by the strong multifractality spectrum \cite{kravtsov_rosen,c_levyloc,c_mblrgeigen,c_mblentropy}.

\subsection{ Inverse Participation Ratios $Y_q$ of arbitrary index $q$}

The corresponding leading behavior of Inverse Participation Ratios of arbitrary index $q$ \cite{mirlin_revue}
\begin{eqnarray}
Y^{(a)}_q \equiv \sum_{l=1}^{N-1} p_l^q \simeq  N \int d\alpha { \cal P}_{a}( \alpha )  N^{-q \alpha} \propto N^{(1-q) D(q)}
\label{yqadef}
\end{eqnarray}
involving the generalized dimensions $D(q)$
is governed by the Legendre transform of the singularity spectrum $f(\alpha)$ as a consequence of the saddle-point evaluation
of the integral.
However here we wish to include also the logarithmic prefactors.

For $a_c=1$, Eq. \ref{facalpha}
yields
\begin{eqnarray}
Y^{(a_c=1)}_q  && \simeq   (\ln N)^2  \int_0^2 d\alpha   \left (1- \frac{\alpha}{2}   \right)  N^{ \left( \frac{1}{2} -q \right) \alpha}
\nonumber \\
&& =  (\ln N)^2  \left[   \frac{ \left( 1- \frac{\alpha}{2} +\frac{1}{2\left( \frac{1}{2} -q \right) \ln N } \right) }{\left( \frac{1}{2} -q \right) \ln N }  N^{ \left( \frac{1}{2} -q \right) \alpha} \right]_{\alpha=0}^{\alpha=2}
\nonumber \\
&& = 
 \frac{ N^{ 1-2q } -1- (1-2q) \ln N   }{ 2 \left( \frac{1}{2} -q \right)^2  } 
\label{yqac}
\end{eqnarray}
so that the leading behavior for large $N$ changes at $q_c=\frac{1}{2}$
\begin{eqnarray}
Y^{(a_c=1)}_{q<\frac{1}{2} } && \oppropto_{N \to +\infty}   \frac{ N^{1-2q} }{ 2 \left( \frac{1}{2} -q \right)^2} 
\nonumber \\
Y^{(a_c=1)}_{q=\frac{1}{2}}  &&  \oppropto_{N \to +\infty}   (\ln N)^2
\nonumber \\
Y^{(a_c=1)}_{q>\frac{1}{2}}  &&  \oppropto_{N \to +\infty}   \frac{\ln N}{\left( q-\frac{1}{2}  \right)} 
\label{yqcriti}
\end{eqnarray}

For $a>1$, Eq. \ref{fabalpha}
yields
\begin{eqnarray}
Y^{(a>1)}_q  \simeq     \frac{ \ln N }{a-1} \int_0^{2a} d\alpha  N^{ \left( \frac{1}{2a } -q \right) \alpha}
 = 
 \frac{ N^{ 1-2a q } -1   }{ (1-a) \left( \frac{1}{2a} -q \right)  } 
\label{yqab}
\end{eqnarray}
so that the leading behavior for large $N$ changes at $q_c=\frac{1}{2a}$
\begin{eqnarray}
Y^{(a>1)}_{q<\frac{1}{2a} } && \oppropto_{N \to +\infty}   \frac{ N^{ 1-2a q }    }{ (1-a) \left( \frac{1}{2a} -q \right)  } 
\nonumber \\
Y^{(a>1)}_{q=\frac{1}{2a}}  &&  \oppropto_{N \to +\infty}  \frac{2a}{a-1} \ln N
\nonumber \\
Y^{(a>1)}_{q>\frac{1}{2a}}  &&  \oppropto_{N \to +\infty}   \frac{ 1   }{ (1-a) \left( q-\frac{1}{2a}  \right)  } 
\label{yqa}
\end{eqnarray}

\subsection{ Discussion }

In summary, the multifractality of eigenvectors for power-law hopping of small amplitude $t$
is the same for quasi-periodic or random energies of finite amplitude :

(i) for $a>1$, the eigenstates remain localized, in the sense that the Inverse Participation Ratios $Y_q$ remain finite above some threshold 
 $q>\frac{1}{2a}$ including $q=1$ corresponding to the normalizability of eigenstates, 
but they are nevertheless described by the multifractal spectrum of Eq. \ref{fab} as a consequence of their power-law localization.

(ii) at the critical point $a_c=1$, the Inverse Participation Ratios $Y_q$ diverge logarithmically above the threshold $q>\frac{1}{2}$,
and the critical eigenstates follow the Strong Multifractality spectrum of Eq. \ref{fac}.

\section{ Perturbation theory in the amplitude $w$ of the Fourier coefficients }

\label{sec_weak}

In this section, we focus on the case of Eq \ref{wb} where the Fourier coefficients $W_m$ of the quasiperiodic potential
 decay as a power-law of exponent $b \geq 1$
with some small amplitude $w$,
while the hoppings $T_l$ of finite amplitude can be either short-ranged or long-ranged.
In particular, the critical case $b_c=1$ of Eq. \ref{wcriti} corresponds in the thermodynamic limit to the $2\pi$-periodic ramp function
of Eq. \ref{wramp}
 with the discontinuity of Eq. \ref{wdiscontinuum}.

\subsection{ First order perturbation theory for the eigenstates }

If $H^{loc}=0$, the $N$  eigenstates are the Fourier modes, 
written here with the quasi-periodic label $Q$ of Eq. \ref{KQ}
\begin{eqnarray}
\vert \psi^{deloc(0)}_{Q} > && = \vert Q >
\label{psideloc0}
\end{eqnarray}
and the corresponding eigenvalues 
\begin{eqnarray}
E^{deloc(0)}_{Q}  && = E^{deloc}_{QQ}
\label{edeloc0}
\end{eqnarray}
are non-degenerate for the critical case $a_c=1$ of Eq. \ref{wcriti} we focus on,
as discussed in more details for the dual case concerning the hoppings (Eqs \ref{edelocdiscrete} \ref{edeloczero} \ref{edelocnearzero}).

The first order perturbation theory in perturbation theory yields the eigenstates 
\begin{eqnarray}
\vert \psi^{deloc(0+1)}_{Q} >&& = \vert Q > + \sum_{Q'} \vert Q' >  \frac{ H^{loc}_{Q'Q} }
{  E^{deloc}_{QQ}- E^{deloc}_{Q'Q'}} 
\nonumber \\
&& = \vert Q > +\sum_{m=-P,+P} 
 \vert Q+m > \frac{ W_m }
{  E^{deloc}_{QQ}- E^{deloc}_{Q+m,Q+m}} 
\label{perweak}
\end{eqnarray}

\subsection{ Inverse participation Ratios $I_q$ in the Fourier basis }

As a consequence, the weights of this eigenstate {\it in this Fourier basis }
\begin{eqnarray}
{\tilde \rho}_m \equiv \left\vert  \frac{ W_m }
{  E^{deloc}_{QQ}- E^{deloc}_{Q+m,Q+m}} \right\vert^2
\label{rhoperweak}
\end{eqnarray}
{\it in this Fourier basis } have exactly the same properties
as the densities {\it in the real space basis} of Eq \ref{pl} for the power-law hoppings discussed in the previous section :
in particular, the Inverse participation ratios  {\it in this Fourier basis }
\begin{eqnarray}
I_q \equiv \sum_m \left\vert  \frac{ W_m }
{  E^{deloc}_{QQ}- E^{deloc}_{Q+m,Q+m}} \right\vert^{2q}
\label{iqfourier}
\end{eqnarray}
follow the equivalent of Eq \ref{yqcriti} for the critical base $b_c=1$
\begin{eqnarray}
I^{(b_c=1)}_{q<\frac{1}{2} } && \oppropto_{N \to +\infty}   \frac{ N^{1-2q} }{ 2 \left( \frac{1}{2} -q \right)^2} 
\nonumber \\
I^{(b_c=1)}_{q=\frac{1}{2}}  &&  \oppropto_{N \to +\infty}   (\ln N)^2
\nonumber \\
I^{(b_c=1)}_{q>\frac{1}{2}}  &&  \oppropto_{N \to +\infty}   \frac{\ln N}{\left( q-\frac{1}{2}  \right)} 
\label{iqcriti}
\end{eqnarray}
and the equivalent of Eq \ref{yqcriti} for $b >1$
\begin{eqnarray}
I^{(b>1)}_{q<\frac{1}{2b} } && \oppropto_{N \to +\infty}   \frac{ N^{ 1-2b q }    }{ (1-b) \left( \frac{1}{2b} -q \right)  } 
\nonumber \\
I^{(b>1)}_{q=\frac{1}{2b}}  &&  \oppropto_{N \to +\infty}  \frac{2b}{b-1} \ln N
\nonumber \\
I^{(b>1)}_{q>\frac{1}{2b}}  &&  \oppropto_{N \to +\infty}   \frac{ 1   }{ (1-b) \left( q-\frac{1}{2b}  \right)  } 
\label{iqb}
\end{eqnarray}

However to analyze the localization properties in real space, one needs instead
to study the Inverse participation Ratios $Y_q$ in the real-space basis.

\subsection{ Inverse participation Ratios $Y_q$ in the real-space basis }

We need to consider the expansion up to second order in the amplitude $w$ to obtain
 the Inverse participation Ratios $Y_q$ in the real-space basis.
Let us assume that the eigenstates are expanded up to second order in the perturbation $H^{loc}$
\begin{eqnarray}
\vert \psi_Q^{(0+1+2)} > = \vert \psi_Q^{(0)} >+ \vert \psi_Q^{(1)} >
+  \vert \psi_Q^{(2)} >
\label{psifourierper}
\end{eqnarray}
The normalization
\begin{eqnarray}
1= <\psi_Q^{(0+1+2)} \vert \psi_Q^{(0+1+2)} > =
\left(  < \psi_Q^{(0)} > \vert +<\psi_Q^{(1)}   \vert +<\psi_Q^{(2)} \vert \right)
\left(  \vert \psi_Q^{(0)} >+ \vert \psi_Q^{(1)} > \right)
+  \vert \psi_Q^{(2)} >
\label{psifourierpernorma}
\end{eqnarray}
yields the following conditions for the first and second orders
\begin{eqnarray}
0 && =<  \psi_Q^{(1)}\vert \psi_Q^{(0)}  >
+<  \psi_Q^{(0)}\vert \psi_Q^{(1)}  > 
\nonumber \\
0 && =<  \psi_Q^{(1)}\vert  \psi_Q^{(1)}  >
+ < \psi_Q^{(2)} \vert  \psi_Q^{(0)}  >
 +<  \psi_Q^{(0)}\vert  \psi_Q^{(2)}  >
\label{densityvan}
\end{eqnarray}

Let us now consider the perturbation up to second order of the  Inverse participation Ratios $Y_q$
\begin{eqnarray}
&& Y_q^{(0+1+2)}  = \sum_{n=1}^N \left( \vert < n \vert \psi_Q^{(0+1+2)}  > \vert^{2} \right)^q
= \sum_{n=1}^N \left(  <  \psi_Q^{(0+1+2)} \vert n > < n \vert \psi_Q^{(0+1+2)}  >  \right)^q
\nonumber \\
&& = \sum_{n=1}^N [  <  \psi_Q^{(0)} \vert n > < n \vert \psi_Q^{(0)}  > 
+ (<  \psi_Q^{(0)} \vert n > < n \vert \psi_Q^{(1)}  >
+ <  \psi_Q^{(1)} \vert n > < n \vert \psi_Q^{(0)}  >)
\nonumber \\
&&+ ( <  \psi_Q^{(0)} \vert n > < n \vert \psi_Q^{(2)}  >
+ <  \psi_Q^{(2)} \vert n > < n \vert \psi_Q^{(0)}  >
+ <  \psi_Q^{(1)} \vert n > < n \vert \psi_Q^{(1)}  >)  ]^q
\label{ipranderson}
\end{eqnarray}

Using the delocalized value of the leading contribution
$<  \psi_Q^{(0)}\vert n  > < n \vert \psi_Q^{(0)}  >  =\frac{1}{N}$, 
the expansion of Eq. \ref{ipranderson} up to second order
becomes
\begin{eqnarray}
&& Y_q^{(0+1+2)}  
 = \sum_{n=1}^N N^{-q} [  1
+ N (<  \psi_Q^{(0)} \vert n > < n \vert \psi_Q^{(1)}  >
+ <  \psi_Q^{(1)} \vert n > < n \vert \psi_Q^{(0)}  >)
\nonumber \\
&&+ N  ( <  \psi_Q^{(0)} \vert n > < n \vert \psi_Q^{(2)}  >
+ <  \psi_Q^{(2)} \vert n > < n \vert \psi_Q^{(0)}  >
+ <  \psi_Q^{(1)} \vert n > < n \vert \psi_Q^{(1)}  >)  ]^q
\nonumber \\
&&= \sum_{n=1}^N N^{-q} [  1
+q  N (<  \psi_Q^{(0)} \vert n > < n \vert \psi_Q^{(1)}  >
+ <  \psi_Q^{(1)} \vert n > < n \vert \psi_Q^{(0)}  >)
\nonumber \\
&& + \frac{q (q-1)}{2}N^2 (<  \psi_Q^{(0)} \vert n > < n \vert \psi_Q^{(1)}  >
+ <  \psi_Q^{(1)} \vert n > < n \vert \psi_Q^{(0)}  >)^2
\nonumber \\
&&+ q N  ( <  \psi_Q^{(0)} \vert n > < n \vert \psi_Q^{(2)}  >
+ <  \psi_Q^{(2)} \vert n > < n \vert \psi_Q^{(0)}  >
+ <  \psi_Q^{(1)} \vert n > < n \vert \psi_Q^{(1)}  >)  ]
\nonumber \\
&&= N^{1-q} 
+q   N^{1-q}     (<  \psi_Q^{(0)}  \vert \psi_Q^{(1)}  >
+ <  \psi_Q^{(1)} \vert  \psi_Q^{(0)}  >)
\nonumber \\
&& +  N^{2-q}    \frac{q (q-1)}{2} \sum_{n=1}^N (<  \psi_Q^{(0)} \vert n > < n \vert \psi_Q^{(1)}  >
+ <  \psi_Q^{(1)} \vert n > < n \vert \psi_Q^{(0)}  >)^2
\nonumber \\
&&+   N^{1-q}   q   ( <  \psi_Q^{(0)}  \vert \psi_Q^{(2)}  >
+ <  \psi_Q^{(2)}  \vert \psi_Q^{(0)}  >
+ <  \psi_Q^{(1)} \vert \psi_Q^{(1)}  >)  
\label{ipranderson1}
\end{eqnarray}

Using Eqs \ref{densityvan}, this simplifies into
\begin{eqnarray}
&& Y_q^{(0+1+2)}  
= N^{1-q} \left[ 
1 +      \frac{q (q-1)}{2}   {\cal S}_N \right]
\label{ipranderson2}
\end{eqnarray}
where
\begin{eqnarray}
{\cal S}_N \equiv  N \sum_{n=1}^N (<  \psi_Q^{(0)} \vert n > < n \vert \psi_Q^{(1)}  >
+ <  \psi_Q^{(1)} \vert n > < n \vert \psi_Q^{(0)}  >)^2  
\label{calsn}
\end{eqnarray}
only depends on the first correction $\vert \psi_Q^{(1)}  > $ of the eigenstate of Eq. \ref{perweak}
\begin{eqnarray}
\vert \psi^{(1)}_{Q} >&& = \sum_{m=-P,+P} 
 \vert Q+m > \frac{ W_m }
{  E^{deloc}_{QQ}- E^{deloc}_{Q+m,Q+m}} 
\label{perweak1}
\end{eqnarray}

Using the explicit expression of Eq \ref{Qfourier} for the Fourier modes,
one obtains
\begin{eqnarray}
<  \psi_Q^{(0)} \vert n > < n \vert \psi_Q^{(1)}  > =\frac{1}{ N}  
\sum_{m=-P,+P}  e^{i 2 \pi m \frac{G n}{N}}
 \frac{ W_m } {  E^{deloc}_{QQ}- E^{deloc}_{Q+m,Q+m}} 
\label{scalQ}
\end{eqnarray}
so that 
\begin{eqnarray}
<  \psi_Q^{(0)} \vert n > < n \vert \psi_Q^{(1)}  >
+ <  \psi_Q^{(1)} \vert n > < n \vert \psi_Q^{(0)}  >
= \frac{1}{ N}  
\sum_{m=-P,+P}  
 \frac{ W_m e^{i 2 \pi m \frac{G n}{N}} +W_m^* e^{-i 2 \pi m \frac{G n}{N}} } {  E^{deloc}_{QQ}- E^{deloc}_{Q+m,Q+m}} 
\label{ipranderson3}
\end{eqnarray}
The sum of Eq. \ref{calsn} thus reads
\begin{eqnarray}
{\cal S}_N &&=  N \sum_{n=1}^N(<  \psi_Q^{(0)} \vert n > < n \vert \psi_Q^{(1)}  >
+ <  \psi_Q^{(1)} \vert n > < n \vert \psi_Q^{(0)}  >)^2
\nonumber \\
&& =
\sum_{m=-P,+P}  \sum_{m'=-P,+P}  
 \frac{  \frac{1}{ N }  \sum_{n=1}^N ( W_m e^{i 2 \pi m \frac{G n}{N}} +W_m^* e^{-i 2 \pi m \frac{G n}{N}} ) 
 ( W_{m'} e^{i 2 \pi m' \frac{G n}{N}} +W_{m'}^* e^{-i 2 \pi m' \frac{G n}{N}} )}
 {  (E^{deloc}_{QQ}- E^{deloc}_{Q+m,Q+m} ) (E^{deloc}_{QQ}- E^{deloc}_{Q+m',Q+m'} )} 
\nonumber \\
&& =
\sum_{m=-P,+P}  \sum_{m'=-P,+P}  
 \frac{ ( W_m W_{m'} +W_m^* W_{m'}^* ) \delta_{m+m'} + (W_m W_{m'}^* +W_m^* W_{m'} ) \delta_{m-m'} }
 {  (E^{deloc}_{QQ}- E^{deloc}_{Q+m,Q+m} ) (E^{deloc}_{QQ}- E^{deloc}_{Q+m',Q+m'} )} 
\nonumber \\
&& =
\sum_{m=-P,+P} 
 \frac{ ( W_m W_{-m} +W_m^* W_{-m}^* )  }
 {  (E^{deloc}_{QQ}- E^{deloc}_{Q+m,Q+m} ) (E^{deloc}_{QQ}- E^{deloc}_{Q-m,Q-m} )} 
+ 
\sum_{m=-P,+P} 
 \frac{  (W_m W_{m}^* +W_m^* W_{m} )  }
 {  (E^{deloc}_{QQ}- E^{deloc}_{Q+m,Q+m} )^2 } 
\label{ipranderson4}
\end{eqnarray}
Since $W_{-m}=W_m^*$, one finally obtains 
\begin{eqnarray}
{\cal S}_N &&=
\sum_{m=-P,+P} 
 \frac{ 2 \vert W_m \vert^2  }
 {  (E^{deloc}_{QQ}- E^{deloc}_{Q+m,Q+m} ) (E^{deloc}_{QQ}- E^{deloc}_{Q-m,Q-m} )} 
+ 
\sum_{m=-P,+P} 
 \frac{  2 \vert W_m \vert^2 }
 {  (E^{deloc}_{QQ}- E^{deloc}_{Q+m,Q+m} )^2 } 
\label{ipranderson5}
\end{eqnarray}
The second contribution is simply $(2 I_1)$ where $I_{q=1}$ of Eq. \ref{iqfourier} is the IPR in the Fourier basis for $q=1$
The first contribution is less singular, since the numerator is the same, but the two denominators are distinct
 instead of coinciding.
So one obtains the following conclusions

(i) for $b>1$, $ I_{q=1}$ of Eq. \ref{iqb} remains finite, so ${\cal S}_N$ remains also finite $O(1)$ in the thermodynamic limit $N \to +\infty$,
and the Inverse Participation ratios keep their delocalized scaling
\begin{eqnarray}
&& Y_q^{(0+1+2)}  
= N^{1-q} \left[ 
1 +      \frac{q (q-1)}{2}   {\cal S}_{N=\infty} \right] \propto N^{(1-q) D^{deloc}(q)}
\label{iprandersonb}
\end{eqnarray}
where the generalized dimensions coinciding with unity
\begin{eqnarray}
 D^{deloc}(q) = 1
\label{dqdeloc}
\end{eqnarray}

(ii) for $b_c=1$, the logarithmic divergence of $I_{q=1}$ of Eq. \ref{iqcriti} induces the logarithmic divergence of the sum
with a small amplitude $(\delta  w^2)$
\begin{eqnarray}
{\cal S}_N   \simeq \delta w^2 \ln N
\label{iprandersonfin}
\end{eqnarray}
As a consequence, the Inverse Participation Ratio
\begin{eqnarray}
&& Y_q^{(0+1+2)}  
= N^{1-q} \left[ 1 +      \frac{q (q-1)}{2}  \delta \ln N \right] \simeq  N^{1-q} N^{\frac{q (q-1)}{2}  \delta }   
= N^{(1-q) D^{criti}(q)} 
\label{iprandersond}
\end{eqnarray}
where the generalized dimensions
\begin{eqnarray}
 D^{criti}(q) = 1- \frac{ q }{2} \delta w^2
\label{iprandersonfindcriti}
\end{eqnarray}
are only sligthly different from their delocalized value (Eq \ref{dqdeloc}).
These properties are well-known as the weak-multifractality regime (see \cite{mirlin_revue} and references therein).

\section{ Conclusion }

\label{sec_conclusion}

In this paper, we have considered the generalized version of
the nearest-neighbor Aubry-Andr\'e quasiperiodic localization model in order to include power-law translation-invariant hoppings $T_l \propto t/l^a$ or power-law Fourier coefficients $W_m \propto w/m^b$ in the quasi-periodic potential. 
We have first recalled the Aubry-Andr\'e duality existing between $T_l$ and $W_m$ when the model is written in the real-space basis and in the Fourier basis on a finite ring. Via the perturbative analysis in the amplitude $t$ of the hoppings, we have obtained that the eigenstates remain power-law localized for $a>1$ and become critical at $a_c=1$ where they follow the Strong Multifractality linear spectrum, as in the equivalent model with random disorder.  Via the perturbative analysis in the amplitude $w$ of the quasi-periodic potential, we have obtained that eigenstates remain delocalized in real space (power-law localized in Fourier space) for $b>1$ and become critical at $b_c=1$ where they follow the Weak Multifractality gaussian spectrum in real space (or Strong Multifractality linear spectrum in the Fourier basis).
 This critical case $b_c=1$ for the Fourier coefficients $W_m$ that we have studied 
corresponds to a periodic linear function with jumps, instead of the cosinus function of the self-dual Aubry-Andr\'e.
More generally, our conclusion is that any periodic function $W(x)$ of weak amplitude $w$ displaying discontinuities, i.e.
characterized by Fourier coefficients decaying only as $1/m$, makes the nearest-neighbor Aubry-Andr\'e model critical.

To go beyond the perturbative analysis in the amplitudes $t$ and $w$ described in the present paper,
 it would be interesting to study numerically
how the multifractal properties of eigenstates evolve as a function of these amplitudes.

\appendix

\section{ Differences between a weak random potential and a weak quasiperiodic potential }

\label{sec_ran}

While the cases of strong random potential and strong quasi-periodic potential are very similar,
the cases of weak random potential and weak quasi-periodic potentials
are completely different as a consequence of the
different couplings existing between the Fourier modes, as discussed in this Appendix.

\subsection{ Generic instability of the weak-disorder expansion around Fourier modes }

For the quasi-periodic case considered in the main text,
the couplings between two Fourier modes $(K,K')$ is governed by Eq. \ref{Hlockkp},
or equivalently by Eq. \ref{Hlockkq} after the relabelling $(Q,Q')$ : the interaction is
thus directly determined by the Fourier coefficients $W_m$.

In the random case where the on-sites energies $H_{nn} $ are random variables of zero-average and variance $w^2$,
the coupling between two Fourier modes $(K,K')$ given by 
\begin{eqnarray}
H^{loc}_{KK'} && = \frac{1}{N} \sum_{n=0}^{N-1} e^{ i 2 \pi (K'-K) \frac{n}{N}}  H_{nn} 
\label{Hlockkran}
\end{eqnarray}
is a random variable of zero-average and of variance 
\begin{eqnarray}
\overline{ \vert H^{loc}(K-K')   \vert^2}= \frac{w^2}{N} 
\label{epsfvar}
\end{eqnarray}
for any difference $(K'-K)$ : effectively, it is thus some 'mean-field' model where all the $N$ Fourier modes are
coupled via some interaction decaying with the system-size $N$ : this is completely different from
the quasi-periodic case considered in the text where the interaction between Fourier modes is governed by
 the Fourier coefficients $W_m$. 
The typical order of magnitude
\begin{eqnarray}
 H^{loc}_N(K-K') \oppropto_{typ} \frac{w}{N^{\frac{1}{2}}} 
\label{epsftyp}
\end{eqnarray}
is much bigger in scaling than the level spacing of the diagonal elements $E^{deloc}_K$ in the Fourier basis
\begin{eqnarray}
\Delta_N \propto \frac{t}{N}
\label{levelspacingn}
\end{eqnarray}
As a consequence, the perturbative expansion of eigenfunctions around Fourier modes
that was described in section \ref{sec_weak} for the quasiperiodic case
looses its meaning in the random case as soon as the ratio
\begin{eqnarray}
\frac{ H^{loc}_N(K-K') }{\Delta_N } \oppropto_{typ} \frac{w N^{\frac{1}{2}}}{t}
\label{ratiow}
\end{eqnarray}
becomes of order unity. The corresponding maximal size
\begin{eqnarray}
N \leq  N_{max} = \left( \frac{t}{w} \right)^2
\label{repsftyp}
\end{eqnarray}
has the same scaling as the localization length $\xi \propto  \left( \frac{t}{W} \right)^2$
that has been computed in the nearest-neighbor Anderson model \cite{kappus,derrida}.

Another way to understand this instability of the weak-disorder expansion 
is that in the Fourier basis, the matrix actually corresponds to the Generalized-Rosenzweig-Porter model 
that has been much studied recently
 \cite{kravtsov_rosen,biroli_rosen,ossipov_rosen,c_levygolden,ioffe} :
 the diagonal elements are finite $O(1)$,
while the off-diagonal elements are random of order $N^{-b}$ : here the value is $b=\frac{1}{2}$ as in the Gaussian Orthogonal Ensemble
and is thus well beyong the critical point $b_c=1$.

\subsection{ Special stability of the weak-disorder expansion for anomalous level spacing }

\label{sec_rod}

The argument above based of the usual level spacing of Eq. \ref{levelspacingn}
has to be modified if the level spacing behaves differently in some region of the spectrum,
as a consequence of some singularity in the density of states.
For instance for the case where the hoppings are real symmetric decaying as the power-law  $T(l)= \frac{t}{\vert l \vert^a}$ 
with $a>1$ \cite{rodriguez,moura,buy,moessner}, the anomalous level spacing near zero momentum $k=0$
\begin{eqnarray}
\Delta_N (k \simeq 0) \propto \frac{t}{N^{a-1}}
\label{levelspacingnano}
\end{eqnarray}
changes the ratio of Eq. \ref{ratiow} into
\begin{eqnarray}
\frac{ H^{loc}_N(K-K') }{\Delta_N } \oppropto_{typ} \frac{W N^{a-\frac{3}{2}}}{t}
\label{ratiowanom}
\end{eqnarray}
so that the weak disorder expansion around Fourier modes is stable for $1<a<\frac{3}{2}$ \cite{rodriguez,moura,buy,moessner}.

\end{document}